# A survey of using EHR as real-world evidence for discovering and validating new drug indications

Nabasmita Talukdar [a], Xiaodan Zhang [b], Shreya Paithankar [b], Hui Wang [e], Bin Chen [a,b,c,d]

[a] Department of Biostatistics & Epidemiology, Michigan State University, East Lansing, Michigan USA

[b] Department of Pediatrics and Human Development, Michigan State University, Grand Rapids, Michigan USA

[c] Department of Pharmacology and Toxicology, Michigan State University, Grand Rapids, Michigan USA

[d] Center for AI-Enabled Drug Discovery, Michigan State University, Grand Rapids, Michigan USA

[e] Lumbrita LLC, Saratoga, California, USA

Corresponding Author: Bin Chen (chenbi12@msu.edu)

## Abstract

Electronic Health Records (EHRs) have been increasingly used as real-world evidence (RWE) to support the discovery and validation of new drug indications. This paper surveys current approaches to EHR-based drug repurposing, covering data sources, processing methodologies, and representation techniques. It discusses study designs and statistical frameworks for evaluating drug efficacy. Key challenges in validation are discussed, with emphasis on the role of large language models (LLMs) and target trial emulation. By synthesizing recent developments and methodological advances, this work provides a foundational resource for researchers aiming to translate real-world data into actionable drug-repurposing evidence.

## Introduction

Drug Discovery is expensive and time-consuming, with average costs amounting to $2-3 billion and requiring 13-15 years of development[1]. In contrast, drug repurposing (or repositioning), discovering new indications for existing drugs, has received considerable attention. Repurposing drugs often have well-studied safety and toxicity profiles, making the process more cost-effective and efficient. Emerging computational and experimental approaches at the preclinical stage have identified numerous potential drug-disease pairs[2,3]. However, translating these into clinical practice requires real-world evidence (RWE) to ensure the repurposed drug's effectiveness and acceptable safety[4].

Although randomized controlled trials (RCTs) are the gold standard for confirming drug efficacy and safety, they are costly, slow, and lack generalizability due to strict criteria and controlled settings[5]. Consequently, RCTs contribute to only 10-20% of clinical decision-making, highlighting the need to utilize RWD as supporting evidence[6]. With the emergence of large-scale Electronic Health Records (EHRs), retrospective validation of repurposing candidates has become increasingly feasible[7]. This is especially important for diseases like Alzheimer's Disease (AD) and COVID-19, where available animal models fail to replicate biology fully. Moreover, many patients present with comorbidities that preclinical models cannot replicate, further underscoring the value of EHRs for drug validation. The recent repurposing of GLP-1 agonists beyond diabetes highlights the role of EHR in identifying new indications[8,9].

While EHR-based drug validation offers major opportunities, it also faces notable challenges. Firstly, EHRs were initially designed for administrative and billing purposes rather than research, leading to variability across institutions in how health records are defined and recorded. This lack of standardization hinders interoperability and complicates the integration and analysis of data across diverse systems. Moreover, many institutes also lack research-ready databases, often requiring significant effort to build custom datasets.

Several factors impact the quality of EHR-based analytics, including data completeness, quality, and cohort size. Low-quality data can introduce biases and reduce the reliability of validation studies. Small patient cohorts limit the statistical power, while unknown confounders may produce false associations. Confirming findings typically requires costly prospective trials, leaving many promising repurposing drugs remain unvalidated. For instance, over 500 drugs have been proposed for AD repurposing in the past decade, yet only 4% have undergone further RWD validation[10].

Despite these challenges, EHRs have unveiled new opportunities to advance the field. The emergence of LLMs and emulation studies also provide new ways for analyzing EHR data and evaluating drug efficacy. This paper reviews commonly used EHR databases, validation techniques, recent advances, and case studies (Fig. 1), with a highlight of the potential of LLMs and emulation in drug repurposing. We aim for this survey to serve as a comprehensive resource for researchers seeking a deeper understanding of computational drug repurposing validation using RWD.

## Literature Search

The literature review followed a structured approach using platforms such as Google Scholar and PubMed. Initial keyword searches (e.g., “Drug Repurposing,” “Observational Studies,” “Real-World Data”) identified relevant papers, followed by a manual review to select those specifically focused on RWD and drug repurposing. Additional studies were sourced through reference lists and iterative searches targeting key aspects of the computational pipeline, including data sources, mining methods, statistical analyses, and selected applications in AD, COVID-19 and cancer. Due to the wide coverage of the field, not all relevant studies could be included.

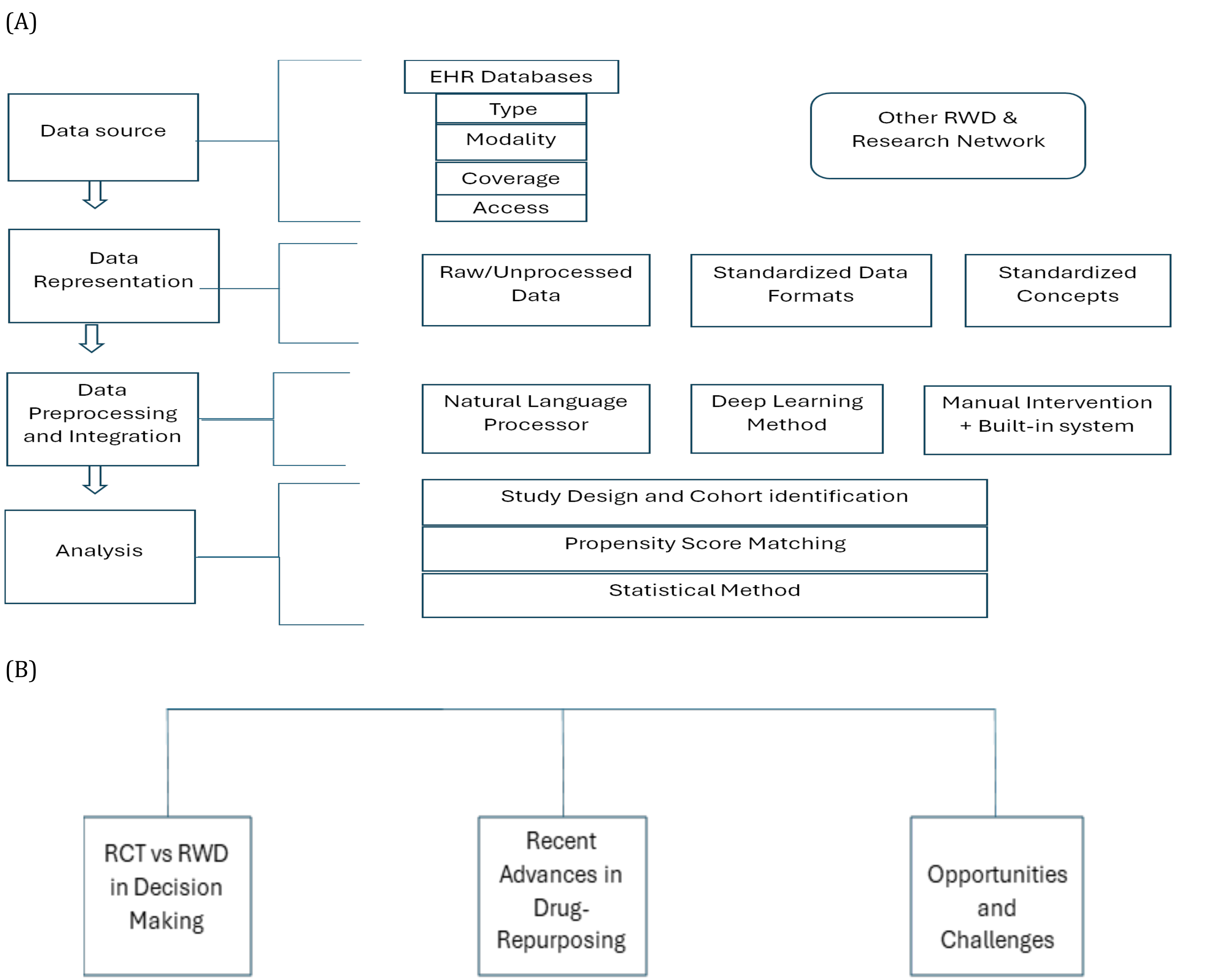

Figure 1: Overview of Survey paper (A) A hierarchical representation of key techniques covered in this paper, outlining the essential steps in drug repurposing using EHR/RWD. (B) Key topics discussed in the paper, include the comparison of RCTs vs emulation trials using RWD for decision-making, recent advancements in drug repurposing, and emerging opportunities and challenges.

## Data Sources

Healthcare providers, such as hospitals, outpatient care centers, rehabilitation centers, insurance companies, laboratories, public health agencies, mental health facilities, and aged care facilities, routinely record patient data digitally during each encounter[11]. EHRs contain a wide range of patient information, including demographics (e.g., age, race, gender, etc.), healthcare encounters, lab results, vital signs, medication prescriptions and dispensing, diagnosis and procedure codes, socio-economic and lifestyle factors, clinical notes (e.g., visit summaries, pathology reports), and imaging data. They also inherently capture temporal data, which is critical for analyzing long-term treatment effects and disease progression. The widespread availability, large sample sizes, and comprehensive socio-demographic data collectively make EHRs a valuable resource for validating drug repurposing candidates.

Table 1 summarized a list of commonly used EHR databases available in the U.S., detailing each database type, modality, population coverage, time span, and accessibility. This comparative analysis enables researchers to identify the most suitable databases for drug repurposing validation. Supplementary Table 1 includes references to example studies that have utilized these databases for validation efforts.

**Table 1: Comparative overview of primary EHR sources for drug repurposing studies available in U.S., listed alphabetically.**

| Database | Type | Modality | Coverage | Access [a] |
|---|---|---|---|---|
| **All of US Research Program dataset** | Consortium Research | EHR, Genomic Data, Surveys, Physical Measurements, Wearable Devices, SDoH | As of April 10, 2025, over 848,000+ participants have enrolled with 470,000+ Electronic Health Records[12]. | Available to approved U.S. researchers |
| **Cerner Real-World Data (CRWD)** | Commercial Research | EHR | 100M+ patient data over 100 healthcare systems across U.S. Initiated in 2019[13]. | Available to U.S. researchers after study approval |
| **EPIC Cosmos** | Consortium Research | EHR, SDoH, Billing Codes, Linked Parent-Child data[14]. | 217M+ Patients records as of 2023, 1,699 Hospitals, launched in 2018[15]. | Restricted to institutional users, and collaborator |
| **Flatiron Health Database[16]** | Commercial | EHR (including Clinical Notes, Oncology-specific Data) | 4 million longitudinal cancer patients from >280 clinics in U.S.[17] | Licensing and partnership |
| | | | | agreement for US researchers |
| **Geisinger Refresh[18]** | Commercial Research | EHR, Claims & Billing | 1.5M+ outpatient data, time coverage of the database spans from 2001 to the present. | Restricted |
| **HealthVerity Database** | Commercial Research | Linked EMR and claims data. | 200M+ patient journeys, 8 years of average data[19]. | Commercially Available |
| **IBM Explorys** | Commercial | EHR, Operational data [b] | 63M+ patient's data from a network of 26 leading healthcare, with an average of 3-4 years per patient. Time coverage – since 1995[20]. | Commercially Available |
| **IBM MarketScan (earlier known as Truven)** | Commercial | Billing, Claims, Carve-out Services, Demographics | Over 250 million patients since 1995. Contains data for patients with continuous insurance enrollment[21]. | Commercially Available |
| **Mayo Clinic Platform_Discover[22]** | Institutional | EHR, Clinical Notes, Imaging, Radiology Reports | 13.6M patients records from Mayo Clinic, including 1.3B+ images, 1.6B lab test results, 10.1M pathology reports, and 698M clinical notes. | Restricted to institutional users |
| **Mount Sinai Data Warehouse Data** | Institutional | EHR, Clinical and Genetic Research Data | 12M+ patient records, most of the data comes from the Epic Clarity and Caboodle databases[23]. | Restricted to institutional users |
| **OneFlorida+ Data Trust** | Consortium Research | EHR, Tumor registry, Death and Claims Data[24] | 17M patient's data, for patients enrolled in Medicaid and from public and private health care systems throughout Florida, Georgia, Alabama[25]. | Restricted to institutional users, and collaborator |
| **Oracle Database** | Commercial | EHR, Billing, Operational, Clinical Notes, SDoH | 108M patients from 100+ healthcare systems across the US, average patient observation period of 7.6 years[26]. | Commercially Available |
| **Premier Health Database[27]** | Commercial Research | Hospital-Based, Service-Level, All-Payer. | 244M+ unique patients, Time coverage from 2000 to present, | Commercially Available |
| | | | 1400 hospitals and healthcare systems. | |
| **The UK Clinical Practice Research Datalink** | Consortium Research | EHR, Lifestyle, Clinical notes | 60M+ patients from over 674 customary practices across the UK, since 1987[28]. | Globally accessible |
| **TriNetX Datasets** | Commercial Research | EHR, Claims, Mortality and Genomic data | 275M+ patients, 70B+ patient specific data on clinical observation spanning around 19 countries[29]. | Licensing and partnership agreement |
| **Truveta Data[30]** | Commercial Research | EHR, SDOH, Claims, Mortality, Clinical Notes, Imaging. | 100M+ patient lives, 8 years of patient history, 5B+ Clinical notes. | Licensing and partnership agreement for US researchers |
| **UCSF (University of California San Francisco) Clinical Data** | Institutional | EHR, Geocoded address data, California Death Registry data, Images, Clinical notes | 6.4M+ patient record. STOR, APeX and Benioff Children's Hospital Oakland data dating back to 1988, 2012 and 2020 respectively[31]. | Restricted to institutional users |
| **Optum Market Clarity data** | Commercial Research | EHR, Claims, Clinic genomics | 100M+ unique patient, 4.5B+ free-text medical notes[32] | Licensing and partnership agreement for US researchers |
| **Vanderbilt University Medical Center (VUMC)** | Institutional | EHR, Genetic Data (BioVU), Clinical notes. | 2.5M+ patients record, within Vanderbilt Health Synthetic Derivative (SD), dating back to the early 1990s, and over 15 years of longitudinal data[33]. | Restricted to institutional users, and collaborator |

a. For more details on the accessibility, refer to Supplementary Table 2
b. The patient journey, provider prescribing patterns, disease progression, procedure volumes, and the overall delivery of care are recorded.

In addition, research networks have also been formed to aggregate EHRs across institutes. For instance, the Patient-Centered Outcomes Research Network (PCORnet) supports clinical research via two Health Plan and nine Clinical Research Networks, with EHR data from 80 million individuals and claims from 60 million (as of 2020)[34]. The Electronic Medical Records and Genomics (eMERGE) network, launched in 2007, integrates DNA biorepositories with EHRs across 10 sites, covering over 105,000 participants[35,36]. The National COVID

Cohort Collaborative (N3C) harmonizes EHR data from multiple U.S. sources, with over 15 million patients, including 5.8 million COVID-19 positive cases (as of July 2022)[37,38]. The Rare Disease Clinical Research Network (RDCRN) focuses on rare diseases, including over 56,000 participants across nearly 200 conditions nationwide[39,40].

Lastly, many databases, though not constituting EHRs, provide valuable information when linked with EHR databases. For example, DrugBank provides detailed molecular and biological information about pharmaceutical drugs. Medication Indication (MEDI) links drugs with their indications and can be used for preliminary information during the validation process or to extract information regarding medication use and their respective indications for all the subjects. SIDER, a public drug side effect database curating 5,868 side effects for 1,430 drugs or therapies, can be combined with EHR databases for comprehensive side effect studies. FDA Adverse Event Reporting System (FAERS) is a post-marketing safety surveillance program for approved drugs containing information related to drug adverse events submitted by manufacturers, healthcare professionals, and the public. The Surveillance, Epidemiology, and End Results (SEER) program provides information on cancer statistics.

## Data Representation and Standardization

Effective data representation and standardization are vital for implementing computational drug repurposing pipelines using EHRs, as healthcare data are collected for varied purposes, causing variations in logical and physical formats. Uniform and accurate clinical data formatting enables robust analysis, integration of diverse sources, and meaningful insights to identify and validate repurposed drug candidates. The process of data representation and standardization can be categorized into three distinct layers.

The first layer consists of unprocessed, raw data from medical records, claims, billing records, patient-reported outcomes (PRO), biobanks or other sources. For instance, the EPIC system is the most widely used system to collect and manage EHR data in health systems. Less expensive systems like Meditech and eClinicalWorks are commonly used in small to medium-sized healthcare systems. REDCap, developed by Vanderbilt University, is a secure, web-based tool for research data collection, scheduling, and surveys. Advanced tools like Clarity, Caboodle (by Epic), and emerging cloud-based platforms like Snowflake support data operations and analysis.

The second layer involves transforming and standardizing data to enable seamless analysis across different platforms and organizations. Standardized data formats facilitate efficient analyses, generate reliable evidence, and reduce inconsistencies. This uniformity is essential for linking and exchanging data between sources, accelerating analysis, and ensuring accurate interpretation. For example, the Observational Medical Outcomes Partnership Common Data Model (OMOP CDM) standardizes the structure and content of observational data and is widely adopted by institutes and consortia like All of Us. FHIR (Fast Healthcare Interoperability Resources) is a standard developed by HL7 (Health Level Seven International) to facilitate the exchange of healthcare information electronically.

The third layer comprises standardized vocabularies representing specific EHR terms consistently across databases (Table 2). These include uniform definitions for conditions, interventions, and outcomes, enabling accurate comparison and integration across data sources to enhance analytical reliability.

**Table 2: Standardized data formats and concepts used in real-world medical data.**

| **Standardized Data Formats** | | |
|---|---|---|
| CPT I, CPT II, CPT III | Current Procedural Terminology | CPT codes provide doctors and healthcare professionals with a standardized language for coding medical services and procedures, aimed at improving reporting accuracy and efficiency. CPT I, CPT II, and CPT III codes are used for reporting medical procedures performed by healthcare professionals, performance measurement and new or developing technology, procedures, and services respectively. |
| CVX | Vaccines Administered | The CVX code set is a standardized set of codes for vaccines administered in the US. |
| DICOM | Digital Imaging and Communications in Medicine | A technical international standard for the digital storage and transmission of medical images and related information. |
| ICD | International Classification of Diseases | For classifying and coding diseases, medical conditions, and other health-related issues. |
| LOINC | Logical Observation Identifiers Names and Codes | A unified Health Informatics standard for representing laboratory test results and vital signs |
| NDC | National Drug Code | A unique 10-digit, 3-segment number that provides information about the drug's manufacturer, product, and package size. |
| NUCC | National Uniform Claim Committee | Provides taxonomy codes which represent healthcare providers based on their specialties, practice areas, and services. |
| OMIM | Online Mendelian Inheritance in Man | Used for describing symptoms, signs, and features of genetic disorders. |
| RxNorm | | A standard nomenclature that provides normalized names and unique identifiers for medicines and drugs. |
| SNOMED CT | Systematized Nomenclature of Medicine-Clinical Terms | A comprehensive clinical terminology system that provides standardized codes for describing and representing clinical concepts such as symptoms, diagnoses, and clinical findings. |
| UCUM | Unified Code for Units of Measure | A coding system for clearly representing measurement units, covering all units currently used in international science, engineering, and business[41]. |
| UDI | The Unique Device Identification | Numeric or alphanumeric code to identify medical devices within the healthcare supply chain. It aims to improve patient safety and facilitate the tracking of medical devices. |
| UMLS | Unified Medical Language System | For integrating and distributing key terminology, classification, and coding standards for multiple kinds of biomedical related concepts. |

## Data Preprocessing and Integration

Data preprocessing and integration are critical steps in the computational drug repurposing pipeline utilizing EHRs, as poor data quality can introduce biases, reduce reliability, and lead to erroneous conclusions. This step involves transforming data into standardized formats, integrating data from multiple sources, and sharing data across units[42]. Using a combination of machine learning techniques, built-in software, manual handling, and tailored programming, structured and unstructured EHR data can be harmonized and mapped to a standard ontology for downstream analysis[2].

Structured data (e.g., age, vital signs) is straightforward to retrieve and process. In contrast, unstructured data (e.g., discharge summaries, and clinical notes) pose challenges due to spelling errors, grammatical inconsistencies, and idiosyncrasies, requiring additional tools. Despite this, the unstructured text contains valuable information and is necessary for capturing complex cases, additional clinical details, and clinical reasoning[43,44]. Extracting meaningful insights from unstructured data typically requires advanced tools and techniques such as Natural Language Processing (NLP) and Machine Learning (ML). Table 3 presents a list of key tools and techniques commonly used for processing medical data in real-world scenarios.

Over the past three decades, several clinical NLP tools have been developed for converting unstructured data into machine-readable format. Numerous studies have shown the effectiveness of NLP in performing various clinical tasks[45,46]. For example, Friedman et al.[47] developed a general natural language text processor for clinical oncology. The Clinical Data Collector (CDC) tool was utilized to extract data on metastatic renal cell carcinoma patients[48].

Deep learning (DL), a subset of ML, has gained significant attention in the field due to its ability to learn abstract representations through deep architectures, often surpassing traditional machine learning approaches[49]. DL has been applied to various healthcare tasks, including phenotype prediction, drug–target interaction identifications, medical imaging analysis, robotic-assisted surgery, and genomics[49,50]. For example, Cheng et al.[51] used DL networks to discover patient phenotypes, and Geraci et al.[52] analyzed unstructured text notes from EHRs to identify youth depression.

NLP-based DL methods have shown improved performance in phenotyping compared to traditional techniques and have been applied in information and relation extraction, and event detection[53,54,55]. Additionally, traditional ML techniques like Naive Bayesian, logistic regression, Random Forest, and Support Vector Machines are also commonly employed for computational phenotype and mining EHR data[11,56]. While this review does not cover these methods in detail, readers may refer to existing comprehensive reviews for deeper insights into ML techniques for EHR data mining[56,57]. More advanced techniques like LLMs are discussed in a subsequent section.

**Table 3: NLP tools and techniques for preparing and transforming medical data in EHR-based drug repurposing**

| Tool/Technique | Name | Description and Applications |
|---|---|---|
| MedLEE | Medical language extraction and encoding system | Extracts information from clinical narrative reports and allows NLP queries. MedLEE was used to identify incidents of adverse events from discharge summaries[58] and colorectal cancer concepts from clinical notes[59]. |
| MedCAT | Medical Concept Annotation Toolkit | An ML algorithm to automatically annotate medical records, using any concept vocabulary including UMLS/SNOMED-CT[60]. |
| NegEx | Negation Extraction | Designed to handle negation cues or negated concepts within clinical text[61]. |
| MedExtractR | | To extract medication information like dosage, frequency, timing, and route of administration[63.] |
| MetaMap | | A tool for extracting biomedical concepts from UMLS in text. |
| MedEx | An open-source existing high-performance NLP system[63] | Medication extraction and normalization tool for clinical text standardization. It was used to extract DM2 drug exposure from unstructured clinical text[65.] |
| cTAKES | Clinical Text Analysis and Knowledge Extraction System | Use to extract clinical data with contextual attributes. Savova et al.[65] applied cTAKES to identify cases of peripheral arterial disease from radiology notes, categorizing them into four groups: positive, negative, probable, and unknown cases[59]. |
| HITEx | Health Information Text Extraction | HITEx is built on top of Gate framework and used for principal diagnoses extraction, discharge medications extraction, and others[66]. |
| CLEF | Clinical eScience Framework[67] | Extraction and identification of semantic entities and relationships. |
| SUTime[68] | | A library for recognizing and normalizing time expressions, which can annotate documents with temporal information. |
| GUTime | | The standard for temporal expression annotation[69], and is part of the General Architecture for Text Engineering framework. |
| MedTime | | An NLP tool for temporal information extraction from clinical text.[70] |
| MLP | Medical Language Processing | This system transforms free-text clinical documents into XML-structured representations. |
| REX | Relation Extraction | To extract various types of relations drug-drug interactions and medication-attribute relations. |
| SymText/ Mplus | A Bayesian Network-based semantic grammar tool | It was developed to extract and normalize findings from radiography reports which is now extended to other uses. Fiszman et al. has applied it to extract the interpretations of lung scans[71] |

## Methods for Temporal Information and Missing Data

Temporal information is extremely valuable for identifying patterns, tracking disease progression, and establishing relationships over time. Sequential models like long short-term memory (LSTM) could be implemented. For example, Baytas et al.[72] introduced a novel time-aware unit designed to learn time decay, addressing irregular time intervals in longitudinal patient records by utilizing LSTM. Lipton et al.[73] evaluated the performance of LSTM in recognizing patterns using multivariate time series clinical data.

Missing values pose challenges due to incomplete entries, follow-up loss, or non-mandatory fields-like lifestyle data-often left blank. Patients may also withhold sensitive information, increasing missingness. Techniques like Last Observation Carried Forward (LOCF), mean/median imputation, Maximum Likelihood Estimation (MLE), ML-based[74] methods (e.g. decision tree, k-nearest neighbors (KNN), and Multivariate Imputation by Chained Equations [MICE]) address missingness. Researchers have found that generative adversarial networks (GANs) can effectively deal with missing data by creating realistic covariates or even creating complete synthetic patients while maintaining cause-and-effect relationships[75,76].

## Data Integration Tools

The availability of built-in functionalities like Application Programming Interfaces (APIs) allows external applications to access and retrieve EHR data[77]. For example, OpenEHR[78], which is an e-health technology with open specifications, clinical models, and software for standardizing clinical data. It is designed to collect, exchange, and store EHR data using an open standard centered around 'archetypes' for detailed clinical information. Heumos et al.[79] recently developed a Python framework "ehrapy", which provides an end-to-end exploratory analysis, including quality control, normalization, feature annotation, and feature summarization.

## The Role of Manual Intervention in EHR Data Preparation

Due to the complex structure of healthcare data, machine learning tools or built-in software often fall short. Creating datasets aligned with research objectives frequently requires manual intervention and custom scripting. Custom variables must often be programmatically created to explore data or uncover patterns. Meticulous data cleaning is vital to fix inconsistencies like duplicate entries, varied units, and non-standard date formats. Transformation may involve generating analysis-specific variables, e.g., age at diagnosis from birthdate, BMI from height and weight, or censoring variables. Domain expertise is critical for accurate data transformation and interpretation. Identifying exposures and outcomes usually requires algorithms to define cohorts and test sensitivity and specificity. These steps are essential for managing generic/brand names, varied units, and inconsistent data entry common in EHRs. Addressing these challenges maintains data integrity and ensures the validity of drug repurposing studies.

## Study Design and Cohort Identification

Selecting an appropriate study design and accurately identifying the cohort based on the research question is crucial in database studies to ensure reliable and valid outcomes. In the context of computational drug repurposing using RWD, all studies are observational and use existing data. Observational studies are further categorized into analytical and descriptive types. Analytical studies are primarily employed to examine the exposure-outcome relationship, whereas descriptive studies focus on summarizing frequency distributions. Analytical studies include three main designs: cohort, case-control, and cross-sectional studies[80] (Fig. 2). Supplementary Figure 1 provides illustrative examples of commonly used designs.

Cohort studies are either prospective or retrospective. In prospective designs, subjects are selected based on their exposure status and followed over time to observe and compare the development of outcomes; baseline cases are excluded. Conversely, retrospective studies, often utilizing EHR/RWD, identify subjects based on past exposure and assess later outcomes. A Cohort study design can be employed to determine if drug exposure improves recovery or lowers mortality, provide temporal relationships, assess multiple outcomes, and ensure high internal validity. Limitations include follow-up loss, selection and information bias, and shifting exposure patterns that may reduce result robustness.

In case-control studies, cohorts are identified by presence/absence of outcome, multiple exposures are examined retrospectively to identify contributing factors. Incident case-control studies involve newly diagnosed cases and matched controls over a study period, while prevalent case-control studies include all cases during a particular timeframe and controls matched to the cases. Ideally, controls come from the same population, when unclear, controls should match cases on demographics and covariates. Case-control studies require fewer subjects than cohort studies and are suitable to study rare and slow-onset diseases. Limitations include inability to provide accurate information on other covariates, susceptibility to recall bias[81], inability to study multiple outcomes, and challenges in choosing appropriate controls.

In cross-sectional studies, outcome and exposure are measured simultaneously. Although cross-sectional studies are valuable for initial explorations in computational drug repurposing, they are less commonly used due to their inability to establish temporal relationships, making it difficult to infer causation between drug exposure and therapeutic outcomes.

(A)

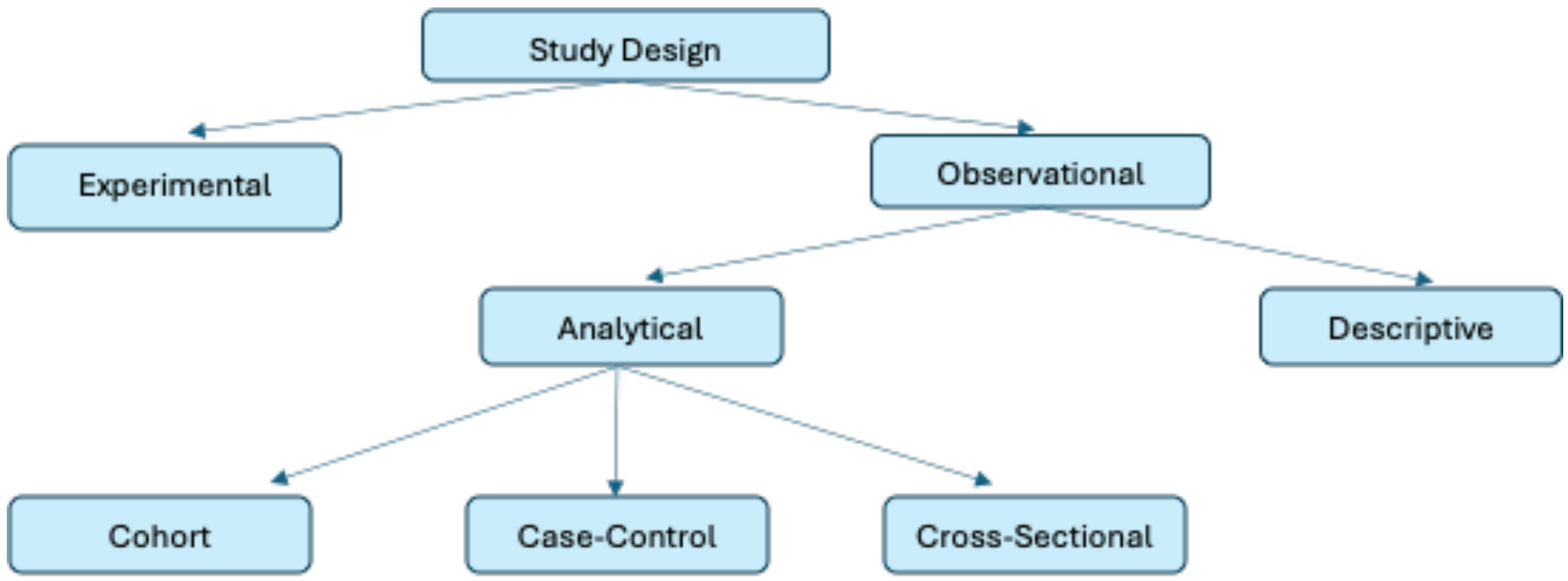
Study Design
Experimental
Observational
Analytical
Descriptive
Cohort
Case-Control
Cross-Sectional

(B)

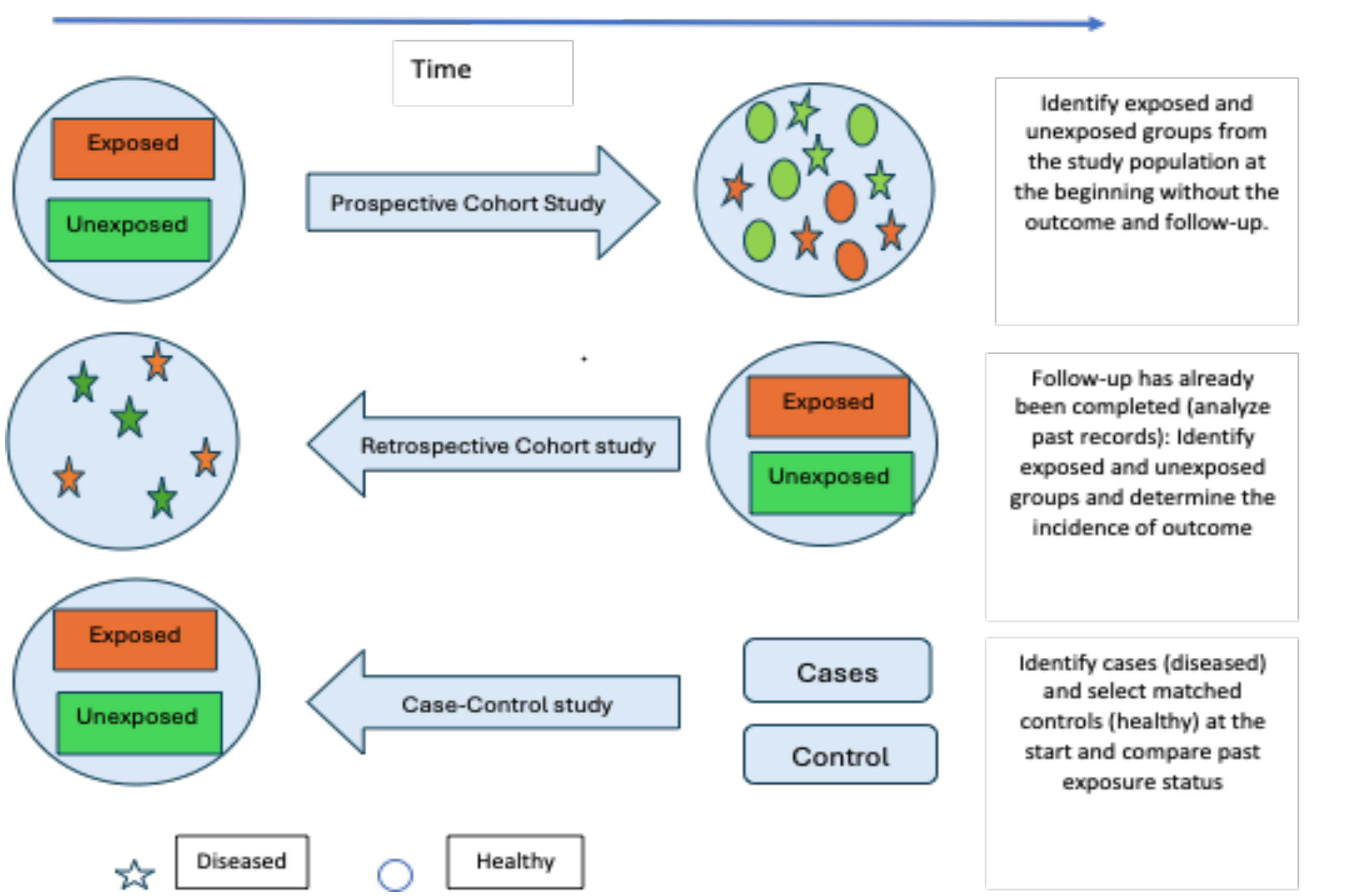
Time
Exposed
Unexposed
Prospective Cohort Study
Identify exposed and unexposed groups from the study population at the beginning without the outcome and follow-up.
Retrospective Cohort study
Exposed
Unexposed
Follow-up has already been completed (analyze past records): Identify exposed and unexposed groups and determine the incidence of outcome
Exposed
Unexposed
Case-Control study
Cases
Control
Identify cases (diseased) and select matched controls (healthy) at the start and compare past exposure status
Diseased
Healthy

Figure 2: (A) Overview of Study Design, and (B) Prospective and Retrospective Cohort Study, and Case-Control Study Design

## Propensity Score Analysis

Accurate measures of the true effect of exposure are challenging due to significant differences in baseline characteristics between groups and the presence of confounding variables. The Propensity Score Analysis (PSA) is a statistical technique used in observational studies to reduce bias between treatment and control groups. The propensity score (PS) represents the probability of receiving a treatment given the observed baseline characteristic[82]. Logistic Regression is the most popularly used technique for estimating PS, and the score could be computed using the following formula.

$$\ln\left(\frac{\mathrm{P(Y=1|X)}}{(1-\mathrm{P(Y=1|X)})}\right) = \beta_0 + \beta_1 X_1 + \cdots + \beta_p X_p$$

$$\mathrm{PS} = \frac{\exp(\beta_0 + \beta_1 X_1 + \cdots + \beta_p X_p)}{(1+\exp(\beta_0 + \beta_1 X_1 + \cdots + \beta_p X_p))}$$

where the exposure is a dependent variable, and the confounding factors are independent variables that may influence treatment assignment.

In addition to traditional logistic regression-based PS estimation, advanced ML and DL models, such as LSTM-PS under the Inverse Probability Weighting (IPW) framework, have shown improved covariate balancing in some cases[83]. Algorithms like Random Forest (RF)[84], Gradient Boosting Machines[85], KNN, and Support Vector Machines (SVMs) offer alternative approaches, each with specific advantages and limitations. For example, RF performs well with missing data, and boosting methods enhance classification accuracy but lack probabilistic outputs. While neural networks show potential for high-dimensional data. SVMs can handle linearly separable classes and non-linear data using kernel functions and do not assume a parametric relationship between the model predictors and outcome[86], while CART is easy to understand and interpret and is suitable for non-linear relationships. However, Zang et al.[87] assessed different propensity score models and discovered that deep learning-based propensity score models did not consistently surpass logistic regression-based approaches in achieving covariate balance. Thus, the exploration of multiple models is recommended.

After computing the propensity scores for each subject in the cohort, common methods to estimate treatment effect include matching, stratification, inverse probability of treatment weighting, and covariate adjustment[88].

Propensity Score Matching (PSM) is the most widely used technique to minimize confounding in observational data. It aims to create a pseudo-randomized dataset by matching case and control groups with similar propensity scores, discarding unmatched subjects. Major methods include greedy and optimal matching. Greedy matching selects the untreated subject with the closest score to a randomly chosen treated subject, repeating until all treated subjects are matched. Optimal matching forms pairs minimizing overall score differences across all pairs[82].

Untreated individuals are selected using methods like nearest-neighbor matching and nearest-neighbor matching within a caliper. In nearest-neighbor matching, the untreated subject with the closest score to the treated individual is paired. If multiple untreated subjects have equal or nearly close propensity scores, a random one is selected. If multiple untreated subjects have similar scores, one is chosen randomly. Caliper matching selects untreated subjects within a defined score threshold. Those without matches are excluded. There is no standard agreed-upon definition of the maximum acceptable distance in caliper matching.

Other methods include stratification, which divides treatment and control groups into PS-based strata and compares average outcomes; covariate adjustment, which includes the propensity score as a predictor in a regression model; and Inverse Probability of Treatment Weighting (IPTW), which assigns weights based on the inverse probability of treatment, balancing covariates and making treatment independent of confounding.

Effective matching also requires variable selection-identifying covariates strongly associated with outcomes but weakly with treatment[89,90]. Causal diagrams can aid in confounding factor selection[91], but challenges arise while dealing with numerous covariates and incomplete knowledge of causal relationships. It is also important to avoid including variables related to the exposure but not the outcome, as they can increase variance without reducing bias[92]. Additionally, this approach necessitates a large sample size, and bias may persist due to unobserved covariates.

In a study investigating the association between semaglutide and suicidal ideation, Wang et al.[8] used 1:1 nearest neighbor greedy matching with a caliper of 0.25 times the standard deviation on covariates that are potential risk factors for suicidal ideation to adjust for confounders in both case and control group. Wang et al.[93] also matched cohorts using 1:1 nearest neighbor greedy matching to study the association of GLP-1 receptor agonists and hepatocellular carcinoma incidence and hepatic decompensation in patients with type 2 diabetes. Laifenfeld et al.[94] emulated clinical trials using observational RWD to identify two drugs, rasagiline and zolpidem as a repurposing drug candidate for progressive diseases like Parkinson's disease (PD). To address potential confounding and selection biases, two distinct causal inference methods are applied: 1) balancing weights through IPW, and 2) outcome models, using standardization.

## Statistical Methods

After the study design and cohort identification, the final step is to assess drug efficacy using statistical methods tailored to the research question. Kaplan-Meier estimate, log-rank test, and Cox proportional hazards model are widely used, especially for survival analysis, to evaluate time-to-event data and compare survival across treatment groups. Additional statistical methods are also used depending on data complexity and study objectives. Table 4 provides a comparative analysis of these statistical methods, highlighting their strengths, limitations and applications, while Supplementary Table 3 provides detailed mathematical formulations.

Method selection depends on factors like research question, study design, and outcome type (e.g., numeric, survival time, or patient counts). For example, Kaplan-Meier estimates survival probability from time-to-event data, whereas Cox models assess covariate effects on hazard rates, allowing multivariate analysis. These techniques enable rigorous evaluation of repurposed drug efficacy, ensuring robust and reliable findings.

**Table 4: Comparative analysis of statistical methods for drug repurposing validation**

| Statistical Method | Strengths | Limitations | Applications |
|---|---|---|---|
| **Kaplan–Meier & Log-rank test**<br><br>Kaplan-Meir is used to evaluate the efficacy of a drug by analyzing survival probability at each time point from patients treated with the drugs, the survival curve is a graphical representation of these probabilities, and Log-Rank (LR) test is a hypothesis test to compare the survival curve of two or more different groups (or drugs). | • Survival probability of all observations is the same.<br>• Handles censored data.<br>• The risk of an event in one group relative to the other remains constant over time.<br>• Distribution of survival time is not required.<br>• LR test is more appropriate to use when the data is skewed. | • Requires sufficient sample size to generate reliable survival curves.<br>• Provides only information about the time of an event of interest, without addressing other important clinical outcomes.<br>• Does not offer an estimate of the magnitude of the difference between groups.<br>• Describes survival based solely on one factor under investigation. | The unadjusted cancer survival probabilities of all exposure groups were visualized using Kaplan–Meier plots in Xu et al.[64]. Kaplan-Meir was used in Rihal et al.[95] to analyze the incidence and prognostic implications of acute renal failure in patients undergoing percutaneous coronary intervention. To minimize the effect from confounding factors, Fang et al.[96] conducted three drug cohort-based observational studies for AD on collected approved drugs and performed unstratified Kaplan-Meier curves and propensity score stratified log-rank tests for each comparison. |
| **Cox proportional-hazards model**<br><br>Evaluate simultaneously the effect of multiple factors on survival and investigate the association between survival time and multiple known variables that potentially affect patient prognosis, adjusting for the effects of confounding factors. | • No functional assumptions are made about the shape of the hazard function.<br>• Can handle multiple covariates (predictors) including both continuous and categorical variables.<br>• The model estimates hazard ratios, which provide the relative risk of the outcome. | • It assumes a linear relationship between the predictor variables and the log hazard ratio.<br>• It assumes that the hazard ratio remains constant throughout the follow-up period.<br>• Independence of survival times is assumed between distinct individuals in the study population.<br>• Couldn't determine whether identified biases have been eliminated[7].<br>• Tendency to overfit, (Vinzamuri et al. introduces methods on handling | Wu et al.[98] proposed a new data-driven approach for drug repurposing that used large EHR database to identify candidate non-cancer drugs and validated them with the same approach in a second EHR database. To assess the relationship between drug exposure and survival, a Cox proportional hazards regression model was used with covariates including patient demographics and tumor information. Similarly, Xu et al.[64] estimated the HRs for all-cause mortalities based on stratified Cox proportional hazard models and evaluated the effects of metformin on survival across multiple cancers. |
| | | correlated features and overfitting[97]). | |
| **Chi-square test**<br><br>Evaluate the independence between different treatment groups, by comparing the frequency of outcome in patients treated with the drug versus those who are not. | • Suitable for categorical data which is commonly encountered in EHR-based drug repurposing validation.<br>• Can handle large sample sizes.<br>• Can analyze multiple outcomes or events. | • Does not provide sufficient statistical power to detect minor differences in small sample sizes.<br>• Does not consider confounding factors.<br>• Subject to the problem of multiple comparisons when analyzing multiple outcomes. | Wang et al.[99] searched public pharmacological databases, genomic databases, and private EHRs for information about drugs and genes associated with glaucoma disease. P-values derived from chi-square tests were used to identify drugs associated with glaucoma-related genes and diseases. |
| **Wilcoxon signed rank test**<br>Use to compare if the difference in two paired groups (before and after drug exposure) is statistically significant.<br>OR<br>**Wilcoxon rank sum test** (Mann-Whitney U test)<br>Use for 2 independent groups<br>OR<br>**Kruskal Wallis test**<br>Use for more than 2 independent groups | • Does not require assumptions about the distribution of the data (Nonparametric test)<br>• Suitable for small sample sizes.<br>• Can be used for both continuous and ordinal data.<br>• Robust for outliers | • Sensitive to ties or data points with the same rank.<br>• Less efficient than parametric tests when the underlying data is normally distributed.<br>• It does not consider the magnitude of the difference. | To validate the drug repurposing candidates, Wu et al.[100] applied the Wilcoxon signed-rank test to assess drug efficacy by comparing the difference in median biomarker measurements taken before and after drug exposure. |
| **Linear mixed model**<br><br>Incorporate fixed and random effects and is used to analyze longitudinal data with continuous repeated measurements from individual subjects over time. | • Useful in hierarchical or longitudinal data structure.<br>• Considers the variability within and between subjects.<br>• Can handle missing data through Maximum Likelihood Estimation. | • Several assumptions need to be held, including normality, linearity, and independence.<br>• Complexity in computation and interpretation | Wu et al.[100] used linear mixed model to assess whether individuals exposed to a repurposing candidate experienced significant biomarker changes.Loucera et al.[101] performed Linear Mixed Effects analysis to determine whether there was an increasing linear trend in the log-transformed lymphocyte progression |
| | • And accounts for repeated measures. | | as a result of receiving treatment for COVID-19. |
| **Logistic regression**<br><br>To model the relationship between a categorical binary dependent variable and one or more independent variables and can be also used for developing predictive models for dichotomous outcomes in medicine[102] | • LR is used in PSA to estimate PS and account for potential confounding.<br>• Can be used in cases of binary outcome. | • Typically requires large sample size.<br>• Requires meeting a few of the assumptions in linear regression.<br>• Proneness to overfitting, which can lead to poor generalizations of the model. | As a method of adjusting for potential confounders, Cheng et al.[103] used multivariate LR to predict the likelihood of receiving treatment versus a comparator based on patient covariates. Shao et al.[104] used an LR model to develop dementia prediction scores. Petrakis et al. used LR to analyze the association between receiving antiviral treatment and clinical outcomes in COVID-19 patients[105]. |
| **Fisher's exact test**<br><br>Use to determine the significance of the association between a drug and outcome for small cohorts. | • Determine the significance of the association.<br>• Use when sample size is small. | • It can only analyze the association between two categorical variables.<br>• Due to its small size the chance of such a sample being representative of a population is low. | In this retrospective case-control study validating predicted drugs with RWD, baseline characteristics of patients on the drugs during testing are compared between case and control groups. The Fisher's exact test is applied to nominal categorical data to assess proportions and calculate p-values[106]. |

## Randomized Controlled Trials vs. Target Trial Emulation Using RWD in Decision Making

Target trial emulation is a methodological approach in epidemiology and clinical research that replicates target randomized trials to estimate the causal effects of interventions using observational data. It mimics the structure and rigor of RCTs, simulating randomization and treatment assignment as if it were an actual RCT. Once eligibility criteria are met and subjects are assigned to trial strategies based on baseline data, outcomes are assessed within a defined period. This method is especially useful for causal inference when a true RCT is not feasible

Though promising for drug repurposing, target trial emulation faces challenges due to RWD's limitations. Caution is needed due to random variation, emulation differences, and biases affecting comparability between RCTs and database studies. A key challenge is defining baseline inclusion/exclusion criteria, often hindered by incomplete histories, inconsistent labs, poorly recorded symptoms, vague investigator criteria, or unidentified confounders. Replicating RCT treatment strategies in RWD is difficult, requiring medication adherence, dosing schedules, and identifying patients on standard care[107]. Emulation also lacks blinding, as patients and providers usually know the treatments[108]. As observational studies are prone to confounding, effective confounding control is crucial. A successful emulation of the target trial's random assignment cannot be achieved if the observational database lacks adequate information on baseline confounders, as it relies on a strong assumption regarding unmeasured confounding[108]. Great care is needed when comparing RCT and RWD findings due to differences in outcome definitions, study populations, selection bias, follow-up lengths, and measured effects. For example, an RCT measuring effect at 2 weeks may not match flexible follow-up in RWD. Emulation discrepancies can reveal whether variations in treatment effect stem from design or randomization differences[109].

Several studies have successfully employed target trial emulation to leverage RWD for drug repurposing. Zang et al.[87] emulated trials for thousands of medications from two large-scale RWD warehouses to identify new indications of approved drugs for AD and perform validation after adjusting for required confounding effects. Hernán et al.[108] introduced the concept of emulating a target trial with big data when a randomized trial is not feasible, emphasizing the importance of evaluating observational data for causal inference. This paper outlined a framework for comparative effectiveness research using big data that explicitly described the target trial and channels existing counterfactual theory[110]. Manuel et al.[111] used a retrospective case-control design with two claims datasets to emulate RCTs assessing montelukast's effect on MS relapses. Chen et al. used RWD to simulate AD clinical trials, results indicated higher serious adverse event rates in the simulated trials compared to the original, demonstrating the feasibility of using real-world data for safety evaluation in Alzheimer's trials[112]. Jeon et al.[113] created an external control arm for COVID-19 patients, showing RWD's potential to replace RCT control arms during emergencies. Daphna et al.[94] identified candidate drugs for PD by emulating phase IIb Clinical Trials from Longitudinal RWD. These findings demonstrate the potential of emulation in validating repurposed drug candidates for conditions lacking disease-modifying treatments.

The role of RWD continues to be debated in drug regulation, health tech assessments, and clinical guidelines[114]. A comparison of 30 completed and 2 ongoing RCTs with insurance claims studies showed strong but imperfect agreement (Pearson correlation = 0.82), with 72% achieving statistical significance[115]. However, cross-sectional review of 220 clinical trials found that only 15% could feasibly be replicated using existing RWD sources[116]. This suggests that while RWD can supplement or replace some clinical trials, observational methods cannot entirely replace traditional clinical trials. A Japanese database replicated three diabetes RCTs and found no agreement with the originals[117]. Collins et al.[118] highlighted RCT strengths and RWD limitations, noting observational studies alone may not reliably ensure effective, safe treatments. Conversely, Raphael et al.[119] presented three successful instances where the FDA used RWE to support regulatory drug approval and highlights that while RWE can aid in conditional drug approvals, RCTs remain essential to confirm treatment efficacy. This work discussed how both can inform decision-making and highlighted that the unique design of RWD, which does not always emulate RCTs, can be seen as a strength because it may better capture the realities of clinical care. It is essential to refine or reject hypotheses based on conflicting results to enhance the validity of both types of studies. Differences between RCT and RWD outcomes, though concerning, are informative and stress the importance of careful emulation and attention to unmeasured confounding. Ultimately, while RWD-based target trial emulation enriches drug repurposing by offering broader insight, RCTs remain critical for internal validity. The integration of RWD and RCTs in decision-making processes offers a balanced approach, leveraging the strengths of both methods while addressing the respective limitations.

## Recent Advances in Drug Repurposing Validation

In the past 5 years, significant progress has been made in identifying and validating repurposed drugs for several diseases using RWD. To provide an overview of these recent advancements, some selected disease-drug pairs for AD, COVID-19 and cancer were showcased in Table 5. The comprehensive list of identified disease-drug pairs using EHR/RWD over the past five years is available in Supplementary Table 4, organized chronologically for each disease category.

We conducted a systematic search on Google Scholar and PubMed for studies over the past five years (2019–2024) with a focus on AD, COVID-19 and cancer. Our search included English-language observational studies, including cohort, case-control, retrospective, and drug-repurposing studies, without geographical restrictions. Specific search terms such as "Disease Name", "Real-World", "Electronic Health Record", "drug repurposing" along with their synonym were employed, using Boolean operators to refine the results as needed. Eligible studies were those that either validated a particular hypothesis or predicted new drug-disease associations. Additionally, we reviewed meta-analyses and systematic reviews to identify relevant retrospective studies and potential drugs showing effects. Studies were excluded if they showed no association, had contradictory responses or negative associations, were prospective studies, used clinical trial data instead of RWD, or focused on other predictive factors or safety rather than efficacy. Each entry in the table includes the disease-drug pair, a summary of the findings, and the corresponding reference.

**Table 5: Highlights of the recent drug repurposing candidates for AD, COVID-19 and cancer, identified over the last 5 years using RWD**

| Disease-Drug Pair | Summary | Ref, Year |
|---|---|---|
| AD - Metformin & Losartan | In this meta-analysis using generative AI and EHR data from VUMC and the All of Us Research Program, metformin and losartan were associated with a reduced risk of AD. Metformin showed a Hazard Ratio (HR) of 0.67 (95% CI: 0.55-0.81), while losartan had an HR of 0.76, (95% CI: 0.6-0.95). | Chao Yan et al.[120], 2024 |
| AD - Pioglitazone & Febuxostats & Atenolol | A pharmacoepidemiology study using the MarketScan Medicare Supplementary database (2012–2017) identified pioglitazone, febuxostat, and atenolol as significantly associated with a reduced AD risk. The HRs were 0.916 (95% CI: 0.861–0.974) for pioglitazone, 0.815 (95% CI: 0.710–0.936) for febuxostat, and 0.949 (95% CI: 0.923–0.976) for atenolol. | Jiansong Fang et al.[121], 2022 |
| AD - Sildenafil | This study developed an AD drug repurposing methodology based on endophenotype disease modules and identified sildenafil as a potential modifier of AD risk. A retrospective case-control analysis of insurance claims for 7.23 million people showed a 69% reduction in AD risk with sildenafil use (HR 0.31, CI 0.25–0.39, $p < 1 \times 10^{-8}$) after 6 years, using Cox regression. | Jiansong Fang et al.[96], 2021 |
| COVID-19 - Nirmatrelvir/Ritonavir or Molnupiravir | A retrospective cohort of COVID-19 patients not requiring supplemental oxygen at admission, treatment with nirmatrelvir–ritonavir (n = 890) or molnupiravir (n = 1856) significantly reduced risks of all-cause mortality and disease progression compared to controls.<br><br>Crude incidence rate per 10,000 person-days for nirmatrelvir–ritonavir is26.47 [CI 21.34-32.46] and for molnupiravir is 38.07 [CI 33.85-42.67]. The HR and CI for nirmatrelvir–ritonavir is (HR 0.34, CI 0.23–0.50, p<0.0001) and for molnupiravir is (HR 0.48, CI 0.40–0.59, p < 0.0001). | Carlos K H Wong et al.[122], 2022 |
| COVID-19 - Melatonin | This study combined network-based prediction with a propensity score matching observational study of 26,779 individuals to explore drug-outcome relationships. Data from the Cleveland Clinic COVID-19 registry showed that melatonin use was linked to a 28% lower likelihood of a positive SARS-CoV-2 test (OR 0.72, CI 0.56–0.91). | Yadi Zhou et al.[106], 2021 |
| COVID-19 - (Multiple Drug) | This study used a deep graph neural network to derive drug representations based on biological interactions, validating them with genetic profiles, in vitro efficacy, and population-level treatment effects.<br><br>Analyzing the Optum de-identified EHR database for COVID-19 patients, six drugs were identified with ATT > 0 and p-value < 0.05: acetaminophen (ATT = 0.25), azithromycin (ATT = 0.18), atorvastatin (ATT = 0.17), albuterol (ATT = 0.14), aspirin (ATT = 0.14), and hydroxychloroquine (ATT = 0.08). | Kanglin Hsieh et al.[123], 2021 |
| Cancer (Epithelial ovarian) - Statin | This large case-control study evaluated the effect of statin use on epithelial ovarian cancer (EOC) risk, analyzing data from 2,040 EOC cases and 2,100 frequency-matched controls in the New England Case Control study.<br><br>Statin users had a 32% lower ovarian cancer risk (OR = 0.68, CI 0.54–0.85), after adjusting for matching factors and covariates. | Babatunde Akinwunmi et al.[124], 2019 |
| Cancer (Endometrial) - Aspirin | A multi-center retrospective study of 1,687 stage I-IV endometrial cancer patients post-hysterectomy showed a 10% improvement in 5-year disease-free survival with low-dose aspirin use (90.6% vs. 80.9%; adjusted HR = 0.46, CI 0.25–0.86), particularly among patients under 60 years old and with a BMI over 30. | Koji Matsuo et al.[125], 2020 |
| Cancer (Liver) - Statin | In this nested case-control study using data from the Scottish Primary Care Clinical Informatics Unit (PCCIU) database, statin use was linked to a lower risk of hepatocellular carcinoma (adjusted HR = 0.48, CI 0.24–0.94). | Kim Tu Tran et at.[126], 2019 |

# Discussion and Perspectives

## Data Access and Privacy Concerns

Access to EHR data is often restricted due to commercial interests, confidentiality agreements, and stringent access criteria, posing challenges for researchers and academic institutions. The lack of publicly available EHR databases or their extremely high costs prevent many researchers from utilizing these valuable resources. Pharmaceutical companies may further regulate access to safety profiles of approved drugs or data on failed or patented drugs, limiting the academic use of data[127]. Moreover, public concerns about using personal health data raise the risk of privacy breaches and unauthorized access, emphasizing the need for robust privacy-preserving measures. Xu et al.[1] proposed privacy-preserving analytics frameworks that bring models directly to the data, avoiding centralized transfers[128]. Techniques such as encryption[129], data de-identification[130], access controls[131], federated learning[132], and audit logs[133] are used to safeguard patient privacy. Legal frameworks like the Health Insurance Portability and Accountability Act (HIPAA) set national standards to protect sensitive patient health information, adding complexity data access and use. Emerging commercial databases such as Truveta offer viable alternatives for researchers lacking institutional access. Initiatives like NIH's All of Us program, which provides researchers with credits to access its database at minimal cost, further prompt equitable access to research resources.

### Data Quality and Integration

EHR data are diverse and heterogeneous, leading to inconsistencies in data representation that complicate integration and timely extraction. While data standardization and data mining tools help address this issue, correcting errors caused by inaccurate entries, software limitations, or poor documentation remains difficult. For instance, time-dependent analyses can be hindered by inaccurate recording of drug start/end dates or disease onset. Often, recorded dates reflect documentation rather than actual events, and medication dispensation or request dates do not guarantee administration. This uncertainty in exposure status reduces analytical reliability. Training and supporting data entry personnel may help to reduce such errors.

Furthermore, health conditions are influenced by factors such as lifestyle, environment, social interactions, or work environment, which are challenging to collect and incorporate but essential for reducing confounding and ensuring a more reliable assessment of drug effects. Prioritizing proper documentation of these factors in EHRs can strengthen drug evaluation. Additionally, challenges in handling missing data, entry errors, and integrating lifestyle or social factors in drug re-purposing validation warrant further exploration.

### Small Sample Size

Small sample size is another common challenge in achieving statistical significance when evaluating drug efficacy. Validating rare disease-drug pairs is often infeasible due to insufficient data. Additionally, small sample sizes hinder the construction of matched control cohorts. To address this, researchers may pool data from multiple sources[134] or apply Bayesian approaches[135], which incorporate prior knowledge and handle small datasets. Relaxing inclusion/exclusion criteria or using meta-analysis, which combines results from multiple studies, can also help increase statistical power.

### Inconsistent Results and Drug-Drug Interactions

While preparing this survey on recent findings in AD, COVID-19 and cancer, we identified several studies with contradictory results[136,137,138,139,140], where some indicated positive effects, others showed no significant association, and some even suggested increased mortality or risk. Such discrepancies raise concerns about the accuracy and reliability of the EHR-based drug repurposing studies. Differences in follow-up periods, study populations, designs, database structures, and statistical methods may contribute to these inconsistent findings. Sharing de-identifiable data and analysis protocols would enhance reproducibility.

Most studies evaluating repurposed drugs did not sufficiently account for drug-drug interactions (DDIs) and side effects, which can lead to unexpected outcomes. Patients often take multiple medications, and interactions among them may compound adverse effects, complicating the assessment of drug efficacy and safety. Yee et al.[141] studied DDIs in COVID-19 treatment, highlighting the importance of rigorous DDI analysis. Gerhart et al.[142] analyzed the top 100 most prescribed drugs in high-risk COVID-19 patients to examine interactions with nirmatrelvir-ritonavir. To further improve the analysis of drug effects, Tatonetti et al.[143] used a data-driven approach to develop two comprehensive databases, one for drug effects and another for DDIs.

Other Applications of EHR in Drug Studies

Recent studies have demonstrated that combining EHRs with other modalities, such as gene expression data, can significantly enhance drug repurposing efforts by corroborating discoveries and revealing mechanistic insights[144]. Zhou et al.[145] developed a network-based prediction model for disease-target interactions by analyzing phenotypic and genetic links between drugs, side effects, diseases, and genes to identify potential drug repositioning candidates for AD.

EHRs also enable the identification of rare or long-term side effects[8], which is critical for evaluating the effectiveness and toxicities of marketed drugs. Side effect data help clarify the correlation between drug targets and adverse events and can support personalized medications. Evaluating efficacy alongside side-effect profiles and comparing repurposed drugs with existing treatments can enhance the robustness of repurposing studies. The FAERS database can be used to compare adverse events between investigational and approved drugs for the same condition. This helps measure the likelihood of adverse events reporting, enhancing the understanding of drug's safety profile. Smith et al.[146] evaluated the risk of adverse drug events for five repurposed COVID-19 regimens in a population of approximately 525,000 Medicare and commercially insured individuals. Moreover, EHRs also support subgroup analyses, offering personalized insights into drug performance across specific patient populations. This underscores the importance of integrating diverse data sources and advanced analytics to ensure the safety and efficacy of repurposed drugs.

## Prospects of Using Large Language Models in Drug Repurposing

LLMs are transforming clinical informatics and medicine[147,148]. Built on the Transformer architecture, these models employ billions, sometimes trillions, of parameters to perform tasks ranging from question answering to complex reasoning. Driven by the scaling laws, which show performance improves with more parameters, companies have rapidly developed LLMs using proprietary datasets and substantial computing resources. While GPT-series models like ChatGPT are the most widely used, their closed-source nature raises significant privacy concerns, particularly in EHR applications. In contrast, open-source models such as DeepSeek, LLaMA, Qwen, and Mistral are evolving rapidly and narrowing the performance gap. Notably, DeepSeek-R1 has achieved performance comparable to GPT-4o in several reasoning tasks, highlighting the feasibility of applying open-source LLMs to EHR data.

The entire field of NLP has evolved rapidly with the release of LLMs. Researchers are now evaluating LLMs for tasks such as summarization[149], reasoning[150], and entity/phenotype recognition[151]. One study discovered that pretraining on clinical tokens enables smaller and more parameter-efficient models to match or outperform larger ones trained on general text[152]. GatorTron trained on >80 billion words of de-identified clinical text, demonstrated strong performance across multiple clinical NLP tasks. Other efforts, like Med-PaLM (a fine-tuned model of PaLM), have shown solid biomedical performance but continue to underperform compared to human clinicians in complex scenarios[153]. GPT-3.5 and GPT-4 have shown promise in semantic similarity and reasoning tasks but remain less effective in information extraction and classification[154]. Ranjani[155] explored LLMs for pediatric Acute Myeloid Leukemia (AML) repurposing and found them useful for structuring clinical trial registry data. Several recent studies have demonstrated that GPT models perform well in clinical information extraction[156,157,158]. For example, Marco et al.[159] assess LLMs for extracting key Social Determinants of Health (SDoH) factors, such as employment status, housing issues, transportation issues, parental status, and social support from clinical data. Liu et al.[160] compared BERT variations for identifying patients with metastatic cancer, finding PubMedBERT fine-tuning most effective. As closed-source models continue to advance, an increasing number of applications are emerging to identify patient cohorts and extract EHR variables for drug validation studies.

Beyond data processing, LLMs show potential for reasoning tasks, including inferring drug-disease associations when trained on EHR data. Studies have shown that GPT-4 performs well in zero-shot settings and remains robust even when keywords or half the input tokens are removed[161]. Using AD as a case study, Yan et al.[120] demonstrated ChatGPT as an AI-driven hypothesis generator for repurposing. Of the drugs suggested, three showed significantly reduced AD risk in real-world datasets. Techniques such as chain-of-thought prompting (CoT), multi-step reasoning, retrieval-augmented generation (RAG), and automated prompt tuning could significantly enhance interpretability, robustness, and performance in complex reasoning tasks[162].

Despite this promise, LLMs raise security and privacy concerns, especially when used with sensitive patient data. Uploading such data to open LLM platforms can risk privacy breaches and unauthorized data access, and the security of closed-source models remains unclear. Robust privacy-preserving analytics and stringent adherence to data protection regulations are essential. Another limitation of LLMs is their reliance on predefined training data, which can pose challenges for underrepresented populations or highly specialized tasks. Additionally, LLMs are prone to issues such as hallucinations, the dissemination of false information, and reasoning errors[163]. Furthermore, EHR databases such as Truveta or All of Us workbenches have proprietary interfaces and external use of LLMs for analysis is currently unsupported, restricting their utility in certain workflows. However, with the feasibility of running open-source models like DeepSeek locally, clinical applications may soon become more practical.

## Declaration of Generative AI and AI-assisted Technologies in the Writing Process

During the preparation of this work, the author(s) used ChatGPT to enhance the readability and language of the manuscript, as well as to ensure compliance with English spelling and grammar. The intellectual content, structure, comparative synthesis of the literature, and all substantive writing were developed entirely by the authors. After utilizing this tool, the author(s) reviewed and edited the content as needed and take(s) full responsibility for the published article's content.

## Declaration of Interest

All authors declare no financial or non-financial conflicts of interest.

## Author Contributions

**N.T.:** Writing – Original draft, Data curation, Investigation. **X.Z., S.P.:** Writing - Review & Editing. **H.W.:** Review. **B.C.:** Writing - Review & Editing, Supervision, Conceptualization.

## Data Availability

No data was used for the research described in the article.

## Acknowledgment

The authors extend their gratitude to Martin Witteveen-Lane, Data Engineer at Corewell Health; Zhehui Luo, PhD, Professor in the Department of Epidemiology and Biostatistics at Michigan State University; and Zeshui(Dominic) Yu, from the University of Pittsburgh School of Pharmacy, for their valuable feedback and suggestions on earlier versions of this manuscript

## Funding

This work is funded by the MSU Strategic Partnership Grants.

**Appendix A.** Supplementary Material

## Glossary

| | |
|---|---|
| AD | Alzheimer's Disease |
| AHR | Adjusted Hazard Ratio |
| ARR | Adjusted Risk Ratio |
| CI | Confidence Interval |
| CRC | Colorectal Cancer |
| DL | Deep Learning |
| DM2 | Type 2 Diabetes Mellitus |
| EHR | Electronic Health Records |
| HR | Hazard Ratio |
| LLM | Large Language Model |
| NHIS | National Health Insurance Service |
| NLP | Natural Language Processing |
| OR | Odds Ratio |
| OS | Overall Survival |
| RCT | Randomized Clinical Trial |
| RWD | Real-World Data |
| RWE | Real World Evidence |
| SDoH | Social Determinants of Health |